\documentclass[10pt]{article}

\usepackage{graphicx}
\usepackage{amsmath}
\usepackage{amssymb}
\usepackage{bm}
\usepackage[utf8]{inputenc}
\usepackage[T1]{fontenc}
\usepackage{makecell}
\usepackage[caption=false]{subfig}

\usepackage{authblk}
\usepackage{geometry}
\usepackage{url}

\newlength{\bibitemsep}
\newlength{\bibparskip}
\let\oldthebibliography\thebibliography
\renewcommand\thebibliography[1]{%
  \oldthebibliography{#1}%
  \setlength{\parskip}{\bibitemsep}%
  \setlength{\itemsep}{\bibparskip}%
}

\title{Spectrally-pure optical serrodyne modulation for continuously-tunable laser offset locking}

\author[1,2,*]{Roame Hildebrand}
\author[1,2]{Wance Wang}
\author[1,2]{Connor Goham}
\author[3]{Alessandro Restelli}
\author[2,4,5]{Joseph W. Britton}

\affil[1]{Institute for Research in Electronics and Applied Physics, University of Maryland, College Park, MD 20742, USA}
\affil[2]{Department of Physics, University of Maryland, College Park, MD 20742, USA}
\affil[3]{Joint Quantum Institute, University of Maryland and the National Institute of Standards and Technology, College Park, MD, USA}
\affil[4]{CCDC Army Research Laboratory, Adelphi, MD, USA}
\affil[5]{Quantum Technology Center, University of Maryland, College Park, MD, USA}
\affil[*]{Email: roame@umd.edu}

\date{}

\begin{document}
\maketitle

\begin{abstract}
The comb-like spectrum added to laser light by an electro-optic modulator (EOM) finds use in a wide range of applications, including coherent optical communication, atomic spectroscopy, and laser frequency and phase stabilization. In some cases a sideband-free optical frequency shift is preferred, such as in laser offset locking using an optical cavity, single-photon frequency shifting, and laser range finding. Approaches to obtaining an optical frequency offset (OFO) involve trade-offs between shift range, conversion gain, and suppression of spurious sidebands. Here we demonstrate an OFO of continuous-wave 871 nm laser light by serrodyne modulation using a fiber EOM and radio-frequency (RF) tones from a commercial RF system on a chip (RFSoC) to achieve shifts of $40$ to $800\text{ MHz}$ with $>15\text{ dB}$ suppression of spurious sidebands and $<1.5\text{ dB}$ conversion loss. We also observe a smoothly varying conversion gain. The utility of this tool is demonstrated by continuously shifting the offset of a cavity-locked laser from $50$ to $1600\text{ MHz}$, a capability useful in spectroscopy of unknown optical transitions.
\end{abstract}

\section{Introduction}
\subsection{OFO Background and Implementations}
The performance requirements for OFO depend on application. Continuously-tunable Pound-Drever-Hall (PDH) laser-offset locks are simplified by spectrally-pure shifts covering ranges greater than half the cavity free spectral range (FSR) and sufficient power stability in the shifted sideband to maintain lock across the sweep bandwidth \cite{drever_laser_1983}. Whereas for single-photon frequency shifting, low insertion loss, high conversion efficiency and precise frequency control is desirable to maximize interference visibility \cite{zhu_spectral_2022, kapoor_electro-optic_2025}. The resolution of laser range finding is limited by optical frequency chirp nonlinearity \cite{uttam_precision_1985,macdonald_frequency_1981}, favoring an OFO with high bandwidth and precise frequency control. A wide range of applications will benefit from improved OFO tools.

Here we show that a commercial arbitrary waveform generator and a waveguide electro-optic modulator enable substantially improved OFO performance for continuous-wave laser light relative to prior demonstrations. We characterize the shift quality via two measures: conversion loss and suppression. Let $P_{\text{in}}$ be the power in the optical fiber upstream of the EOM and $P_{\text{out}}$ be the power out of the EOM with no RF applied. Conversion loss is defined as the ratio $\beta=P_{\text{out}}/P_{\text{shift}}$ and suppression as the ratio $\gamma=P_{\text{shift}}/P_{\text{spur}}$, where $P_{\text{shift}}$ is power at the target frequency in the output spectrum and $P_{\text{spur}}$ is the maximum power at any frequency that is not the target frequency in the output spectrum. We also define insertion loss $\alpha=P_{\text{in}}/P_{\text{out}}$.

The most straightforward and flexible method of obtaining light at a frequency offset is beat-note locking a pair of independent lasers. This approach admits a wide tuning range, fast frequency shifting (limited by laser lock bandwidth), high optical power, and zero sidebands, but requires the use of a second laser. The most general but expensive approach is offsetting a pair of lasers via locking to an optical frequency comb. Free-space acousto-optic modulators (AOMs) rely on phonon scattering and permit single-pass shifts of $f_{a}$ where $80\text{ MHz}<f_{a}<500\text{ MHz}$, high optical power handling, and excellent carrier suppression, but suffer from low tuning bandwidth ($\lesssim\frac{1}{5}f_{a}$), moderate insertion loss ($\sim1\text{ dB}$) and a frequency-dependent angular shift. Multi-pass AOM setups \cite{donley_double-pass_2005} permit a greater absolute shift and remedy the frequency-dependent angular shift but have limited tuning bandwidth, high insertion loss (e.g. $5\text{ GHz }\pm150\text{ MHz}$, $-10\text{ dB}$ \cite{zhou_laser_2020}) and are challenging to keep aligned. Electro-optic phase modulators (EOMs) rely on the electro-optic effect to mix optical and microwave signals. They are an attractive option for larger frequency shifts with the added complexity of polarization sensitivity. Free-space EOMs offer high optical power handling but often rely on high-quality-factor resonators to obtain the necessary RF voltage, which constrains the RF bandwidth. Waveguide-based EOMs have greater bandwidth ($\sim50\text{ GHz}$) and high modulation efficiency but low optical power handling. 

There are several EOM-based approaches to OFO. A single optical sideband can be obtained from interference of EOM-generated in-phase and in-quadrature (IQ) optical fields, but the minimum conversion loss is $4.7\text{ dB}$ and the technical overhead is high \cite{kawanishi_high-speed_2007,tu_quadrature_2024,kodigala_high-performance_2024,lo_precise_2017}. For example, Kawanishi et al. used 4 EOMs, 4 DC biases, and 4 RF drives to obtain shifts of $\pm40\text{ GHz}$ with $6.9 \text{ dB}$ insertion loss and $17 \text{ dB}$ suppression. In 2021, the Loncar group used a pair of evanescently coupled lithium-niobate ring resonators driven by an RF tone to implement on-chip OFO with a conversion loss of $<1.2\text{ dB}$ from $8\text{ GHz}$ to $11\text{ GHz}$ \cite{hu_-chip_2021}. This novel approach may be the future of OFO in highly integrated devices, but is not yet commercially available.

\subsection{Serrodyne modulation}
Serrodyne modulation is an alternative approach that applies a quasi-continuous optical phase ramp to generate an OFO. A voltage $V(t)$ applied across the electro-optic material shifts the optical phase: $E_{0}\sin\left(2\pi\xi t+\phi(t)\right)$, where $\xi$  is the input optical frequency and $\phi(t)=\frac{\pi}{V_{\pi}}V(t)$. For serrodyne modulation the voltage waveform is a sawtooth 
\begin{equation}\label{eq:Vserrodyne}
    V_{s}(t)=2NV_{\pi}\left[\left(f_{m}t\enspace\text{mod}\enspace1\right)-\frac{1}{2}\right]
\end{equation}
with frequency $f_{m}$ and peak-to-peak amplitude $2NV_{\pi}$ for $N\in\mathbb{N}$. The idealized EOM output is then $\left(-1\right)^{N}E_{0}\sin\left[2\pi(\xi+Nf_{m})t\right]$, effectively shifting the laser frequency by $Nf_{m}$. In practice, the output has unwanted spectral features due to imperfections in $V_{s}(t)$, improper impedance matching to the electro-optic medium, and RF amplifier finite bandwidth and frequency-dependent gain.

An early effort by Johnson and Cox (1988) obtained a serrodyne shift of $5\text{ kHz}$ using a voltage ramp produced by charging a capacitor \cite{johnson_serrodyne_1988}. A later innovation exploited the extreme nonlinearity of commercially-available nonlinear transmission lines (NLTL) \cite{houtz_wideband_2009,johnson_broadband_2010}. For example, the Kasevich group obtained a conversion loss of $1.0\text{ dB}$ at $1.5\text{ GHz}$ shift, $10\text{ dB}$ suppression over $940$ to $1200\text{ MHz}$, and $7\text{ dB}$ suppression over $200$ to $1240\text{ MHz}$ \cite{johnson_broadband_2010}. Wide bandwidth OFO proved challenging with NLTLs despite experimentation with a range of devices. All-optical serrodyne modulation was demonstrated using cross-phase modulation in a nonlinear fiber but has high technical overhead and small bandwidth \cite{lee_all-optical_2021}. Recently, modern arbitrary waveform generators (AWG) have enabled improved sawtooth waveforms at low cost. Minjeong Kim et al. obtained conversion losses as low as $0.13\text{ dB}$ over tens of megahertz, but did not attempt modulations beyond $400 \text{ MHz}$ \cite{kim_40_2020}. Table \ref{tab:Summary-of-OFO} compares the range of approaches.

\setlength{\tabcolsep}{4pt}
\begin{table*}[h]
    \caption{
        \protect\label{tab:Summary-of-OFO}Comparison of OFO methods for CW light with an emphasis on high bandwidth and spurious signal suppression. SM denotes serrodyne modulation, NLTL denotes nonlinear transmission line, OC denotes on-chip, and X indicates unreported data. The wavelength column indicates the input laser color used for the demonstration. The columns include the widest frequency range over which OFO demonstrations exhibited a particular level of conversion or suppression. Some demonstrations discussed in this paper are not included due to lack of broadband ($>100\text{ MHz}$) testing. Relative to prior work, we improve on spur suppression bandwidth by a factor of 5.
    }
    \vspace{0.2cm}
    \hspace{-2.5cm}
    \def\arraystretch{1.5} 
    \begin{tabular}{|c|c|c|c|c|c|c|}
        \hline
        Source & OFO Method & \makecell[t]{Wavelength\\(nm)} & \makecell[t]{Shift Range\\(MHz)} & \makecell[t]{$<1\text{dB}$\\Conversion Loss\\(MHz)} & \makecell[t]{$<2\text{dB}$\\Conversion Loss\\(MHz)} & \makecell[t]{$>10\text{dB}$\\Suppression\\(MHz)\textsuperscript{\textdagger}}

        \\
        \hline
        \cite{hu_-chip_2021} & coherent waveguide coupling (OC) & 1,601 & 8000-11,000 & 8140-11,000 & 8000-11,000 & X\\
        \cite{houtz_wideband_2009} & SM via NLTL \& EOM & 850 & 100-1650 & 1490-1500 & 960-1070 & X\\
        \cite{johnson_broadband_2010} & SM via NLTL \& EOM & 780 & 200-1600 & 500-530 & 670-1240 & 940-1200\\
        \cite{barbiero_broadband_2023} & SM via NLTL \& EOM & 698 & 50-750 & None & 610-640 & 600-650\\
        \cite{kim_40_2020} & SM via AWG \& EOM & 780 & 3, 28* & X & X & X\\
        Here & SM via AWG \& EOM & 871 & 40-1600 & 40-550 & 40-1330 & 40-1490\\
        \hline
    \end{tabular}
    \vspace{0.1cm}
    \begin{flushleft}
    \hangindent=0.2cm
    \hangafter=1
    *Broadband performance not demonstrated. Achieved $<0.15\text{ dB}$ conversion loss and $>26\text{ dB}$ suppression for reported shift frequencies.\\
    \textsuperscript{\textdagger}Suppression metric includes optical carrier.
    \end{flushleft}
\end{table*}

Here we show that the use of low-cost, high-bandwidth AWGs permits substantial improvement in high-bandwidth OFO using off-the-shelf components. In this scheme, an AWG provides the voltage ramp, which is then amplified and applied to a fiber EOM. We demonstrate the utility of our OFO by using it to continuously tune the offset of an 871 nm laser from an optical cavity mode by 50 MHz to 1600 MHz.

\section{Apparatus}

\begin{figure}
    \centering
    \includegraphics[width=0.85\linewidth]{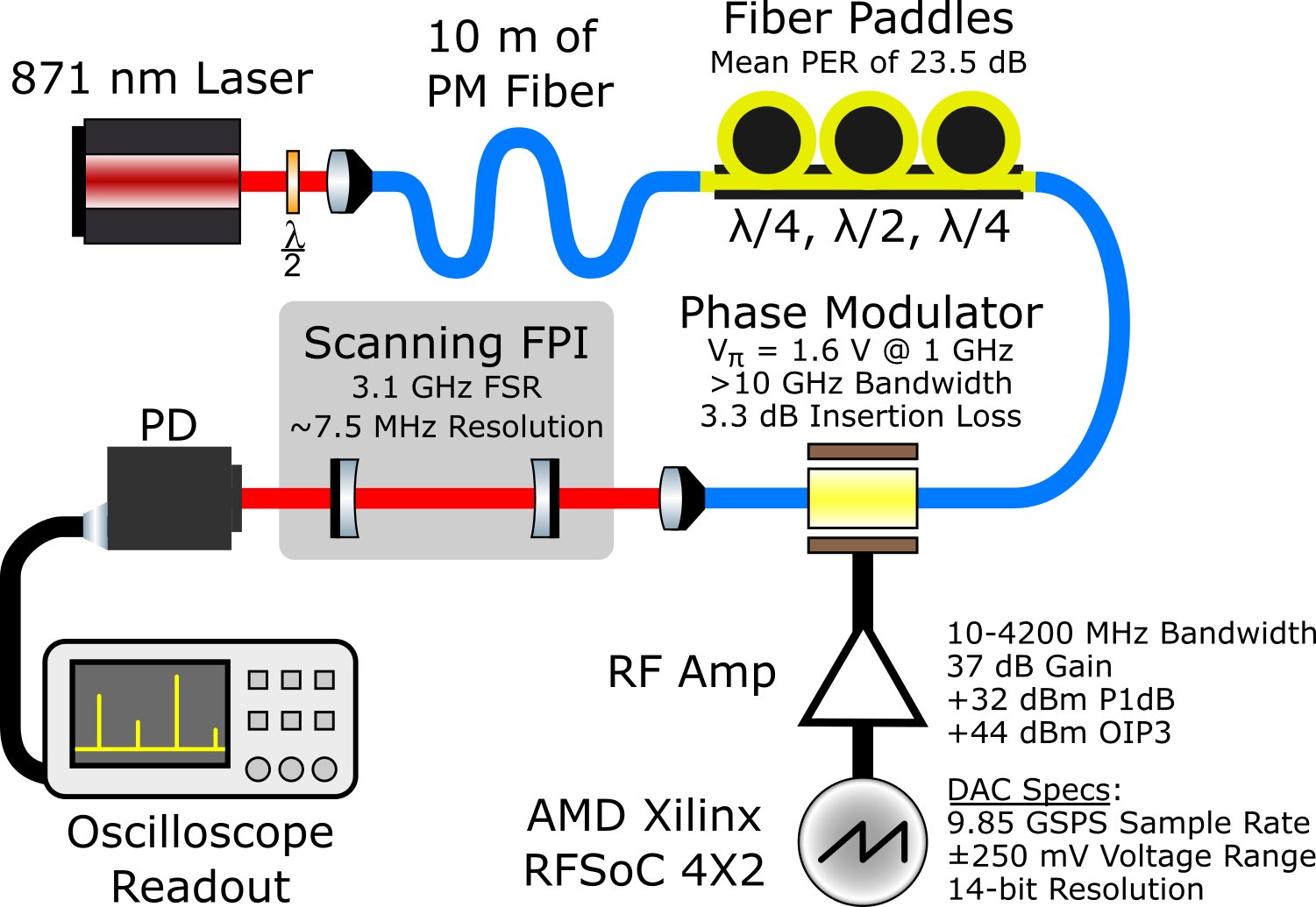}
    \caption{\label{fig:serrodyne-setup} Experimental apparatus used for evaluating the phase-modulated spectrum. Blue lines indicate PM Fiber, yellow lines indicate non-PM fiber, red lines indicate free-space beam propagation, black lines indicate electrical connections.}
\end{figure}

Our experimental apparatus is shown in Figure \ref{fig:serrodyne-setup}. The optical elements of our apparatus consist of a laser, fiber polarization paddles, a fiber EOM, and an optical spectrum analyzer. The single-mode laser is an 871 nm ECDL (Toptica DLC Pro). The EOM is a LiNbO3 phase modulator with $3.3\text{ dB}$ insertion loss, $27\text{ dBm}$ max RF input power, $V_{\pi}=1.6\text{ \mbox{V}}$ (measured at $1\text{ GHz}$), and an electro-optic modulation response that varies by up to $3\text{ dB}$ over DC to $14\text{ GHz}$ (EOSPACE PM-0S5-10-PFA-PFA-800-P5). The EOM is packaged with FC/APC connectors and Panda-type polarization maintaining (PM) fiber. We used polarization paddles ($\lambda/4$, $\lambda/2$, $\lambda/4$) to align the polarization with the EOM modulation axis and obtain an average polarization extinction ratio (PER) of $23.5\text{ dB}$ at the input of the EOM fiber.

The optical spectrum analyzer is a scanning Fabry–Pérot interferometer (FPI) with $\sim7.5\text{ MHz}$ resolution and $3\text{ GHz}$ FSR (Thorlabs SA200-8B). We measure an optical power spectrum by scanning the FPI cavity length across the cavity FSR while monitoring transmission on a photodetector. We confirmed that we are operating the FPI photodetector in the linear regime. Our measurements are insensitive to nonlinearity and hysteresis in the FPI PZT as we are only measuring cavity transmission of discrete spectral features, not their relative positions. We only explored OFO $>40\text{ MHz}$ due to limited FPI resolution.

\subsection{AMD Xilinx RFSoC 4x2 configuration}
The RF serrodyne signal is generated by an AWG followed by an amplifier, which then drives the EOM. The RF amplifier has a bandwidth of $10$ to $4200\text{ MHz}$, $37\text{ dB}$ gain varying by at most $1.5\text{ dB}$ over its bandwidth, $+32\text{ dBm}$ P1dB, and $+44\text{ dBm}$ OIP3 (Mini-Circuits ZHL-10M4G21W0+). There are many commercially available AWGs capable of exceedingly high sampling rates, well over 100 GSPS. Although we expect these devices can achieve remarkable serrodyne performance, we chose the AMD Xilinx RFSoC 4x2 as an economic alternative that could still yield competitive results. This device has two DACs capable of 10 GSPS for the modest cost of \$2,149. This, in addition to the high bandwidth amplifier, brings the cost of the modulation signal synthesis system to \$3,808.

The AMD Xilinx RFSoC 4x2 is a development board based on the AMD ZYNQ Ultrascale+ RFSoC ZU48DR device. The AMD device is a monolithic radio-frequency system on chip (RFSoC) that contains a processing system (PS) with a quad-core 64-bit ARM processor and a dual-core 32-bit ARM, a field programmable gate array (FPGA) with nearly a million programmable logic cells (PL), and eight 14-bit digital to analog converters (DAC) with a maximum sampling rate of $10\text{ GSPS}$ and voltage range of $\pm250 \text{ mV}$. The RFSoC 4x2 development board exposes only 4 ADC channels and 2 DAC channels to the user.

We wrote in Verilog a ramp generator core based on a 32-bit accumulator with 32-bit programmable increment \textit{inc} to feed one of the RFSoC DAC channels with a digital ramp. Since the SoC PL cannot work at the $10\text{ GHz}$ rate of the DAC channel, we enabled multiplexing in the DAC and the ramp generator core is run at a clock frequency of $615\text{ MHz}$ and replicated 16 times to calculate 16 output samples for each PL clock cycle. The resulting 32-bit registers are multiplied by a 16-bit gain factor \textit{g} and the 14 most significant bits of the 16 48-bit results are fed to the DAC core during each PL clock cycle resulting in a $9.85\text{ GSPS}$ analog output. The programmable increment \textit{inc} and gain \textit{g} are memory mapped to the PS memory using automatically generated cores available in AMD Vivado.

The PS runs PYNQ (Python Productivity for ZYNQ), a Linux Ubuntu-based distribution, and allows the user to write Python application programming interfaces (API) to interface PL and PS through memory mapped registers or direct memory access (DMA). In our specific case we wrote an API to adjust amplitude and frequency of the sawtooth waveform. The Python code also manages the conversion between physical units (normalized scaling factor and Hz) and hardware units (16 and 32-bit codes) used in \textit{g} and \textit{inc} registers. The latency for changing the shift frequency via python executed on the board in the current unoptimized implementation is $\sim75\;\mu\text{s}$–the minimum possible latency is $40\text{ ns}$ set by the DAC pipeline. The source for the ramp generator is available on GitHub: \url{https://github.com/JQIamo/ramp_generator_RFSoC4x2}.

\begin{figure}
    \centering
    \hspace*{-1.5cm}
    \subfloat[50 MHz modulation frequency.]{\includegraphics[width=0.6\textwidth]{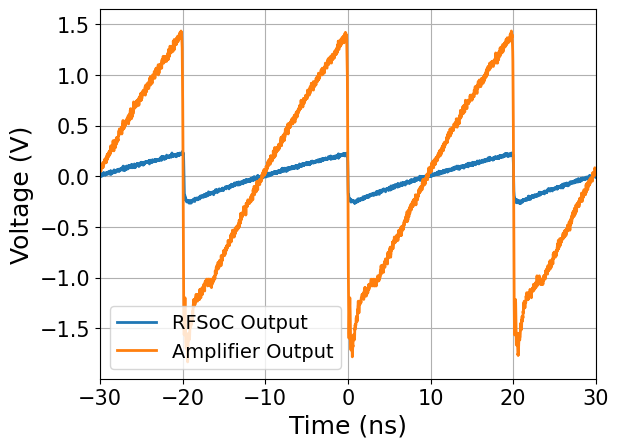}}
    \hfill
    \subfloat[1600 MHz modulation frequency.]{\includegraphics[width=0.6\textwidth]{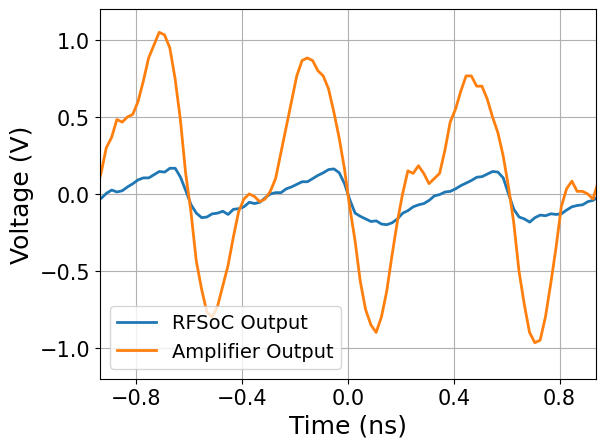}}
    \hspace*{-1.5cm}

    \caption{\protect\label{fig:Serrodyne-example_traces}Example traces of the synthesized serrodyne signal before and after the amplifier. Both signals are sampled at 50 GSPS here. The amplitude shown is slightly less than what would minimize the conversion loss at each frequency.}
\end{figure}

\begin{figure}
    \centering
    \includegraphics[width=0.6\textwidth]{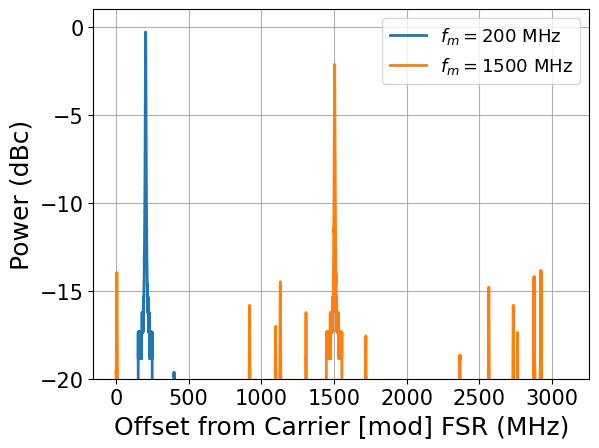}
    
    \caption{\protect\label{fig:example-spectra} Example post-modulation spectra as observed on our optical spectrum analyzer.}
\end{figure}

\section{Measurements}

We characterized our OFO by measuring its performance at shift frequencies spaced by $20 \text{ MHz}$ over the operational range of the sawtooth generator and spectrum analyzer. At each frequency we vary the amplitude of the waveform until optical power in the first sideband is maximized. We then measure the amplitude of the carrier, the first sideband, and the most prominent spurious feature. Then, with RF off, we measure the carrier. This is repeated 5 times for each frequency and the average performance is recorded. We observed that the PER of the fiber link that transmits laser light to the EOM drifts by $\pm2.6\text{ dB}$ over an hour, so we frequently check the PER and adjust the paddles upstream of the EOM as necessary. Ultimately, PER limits the maximum contrast of our measurements to $26\text{ dB}$.

\begin{figure}
    \centering
    \hspace*{-1.5cm}
    \subfloat[Conversion loss for $N=1$.]{\includegraphics[width=0.6\textwidth]{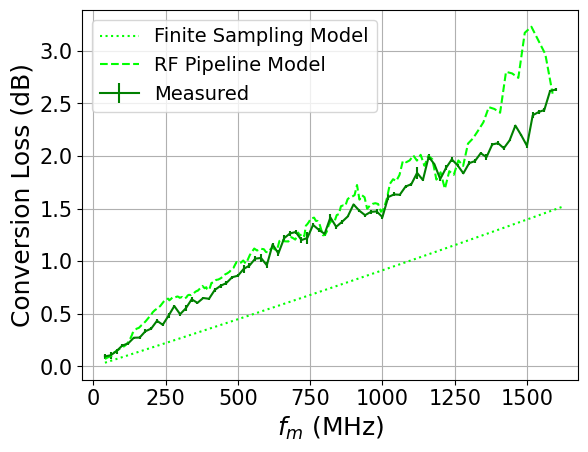}}
    \hfill
    \subfloat[Suppression for $N=1$.]{\includegraphics[width=0.6\textwidth]{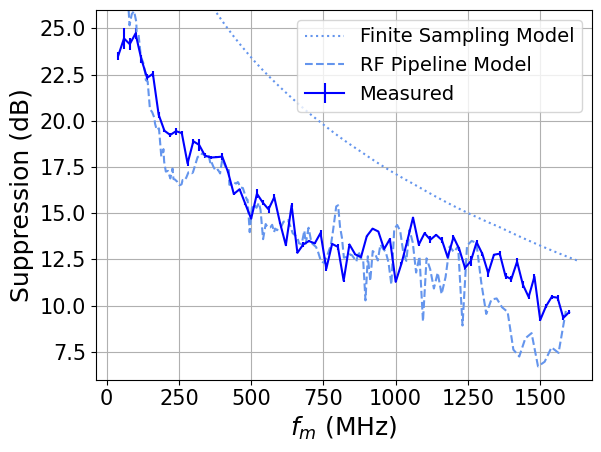}}
    \hspace*{-1.5cm}

    \caption{\protect\label{fig:OFO-combined-performance}OFO performance for N=1 as a function of the modulation frequency.}
\end{figure}

Example traces of the scanning Fabry-P\'erot readout are shown in Figure \ref{fig:example-spectra} and our observations for $N=1$ are in Figure \ref{fig:OFO-combined-performance}. A conversion loss of $<1\text{ dB}$ was maintained over $40-550\text{ MHz}$, $<2\text{ dB}$ over $40-1330\text{ MHz}$ and a suppression of $>10\text{ dB}$ maintained over $40-1490\text{ MHz}$. The lowest conversion loss was $0.095\pm0.020\text{ dB}$ at $40\text{ MHz}$ and the highest suppression was $24.72\pm0.28$\text{ dB} at $100\text{ MHz}$. The broadband conversion loss was smoothly-varying, while suppression showed greater irregularities (Figure \ref{fig:OFO-combined-performance}). It can be shown that the magnitude of the spurious sideband electric fields are sensitive to errors in the amplitude of the corresponding harmonic of the modulation signal to first order. This fact, paired with the varying frequency response of our RF pipeline (illustrated in Figure \ref{fig:Measured-RF-to-optical-transfer}) yields erratic suppression performance with frequency as shown here.

\subsection{Numerical modeling}
To better understand our observations we explored two models. The first model we call the \textit{Finite Sampling model}, which gauges the impact of finite AWG sampling frequency $f_{s}$. This model samples an ideal sawtooth wave of frequency $f_{m}$ at a sampling rate $f_{s}$ and interpolates between these samples to produce the modulation waveform. We assume the DAC analog front end permits full-scale voltage swings in the interval $1/f_{s}$. This model differs from an ideal sawtooth in two ways: there is a nonzero reset time between consecutive ramps, set by the period between samples, and a reset may begin just before or just after having ramped $2\pi$ in phase. The former feature is the primary limiter on performance here. To achieve a conversion loss $<1\text{ dB}$ or suppression $>15\text{ dB}$, $f_{m}/f_{s}<0.11$ is required. The dotted lines in Figure \ref{fig:OFO-combined-performance} show the calculated performance of the Finite Sampling model. It's clear from measured performance that finite AWG sampling frequency didn't limit our OFO. 

\begin{figure}
    \centering
    \includegraphics[width=0.6\textwidth]{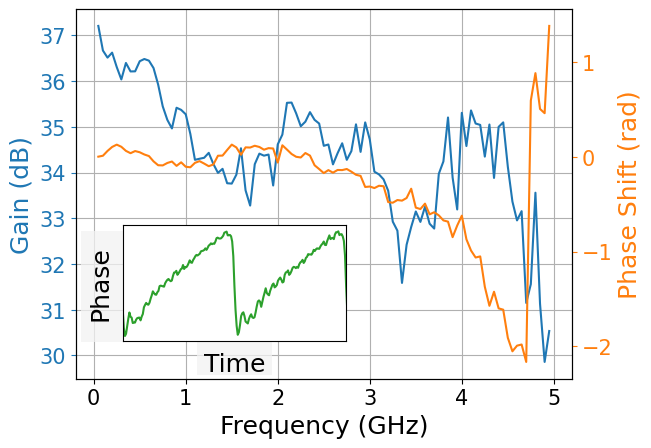}
    \caption{\protect\label{fig:Measured-RF-to-optical-transfer}Measured DAC-to-optical transfer function $T_{\text{DTO}}$. To illustrate the the effect of this nonuniform transfer function, the plot in the bottom left shows the resulting phase modulation for a sampling-frequency-limited input saw wave at $300\text{ MHz}.$}
\end{figure}

Our second model, which we call the \textit{RF Pipeline model}, considers the contribution of RF pipeline imperfections. We used a $50\text{ GSPS}$ oscilloscope (Tektronix DSA72004B) to characterize the transfer function of the amplifier. We combined this with manufacturer's factory test data for the EOM to estimate the overall DAC-to-optical transfer function $T_{\text{DTO}}$ (shown in Figure \ref{fig:Measured-RF-to-optical-transfer}), which maps the signal present at the DAC output (in the frequency domain) to the induced phase modulation at the EOM (in the frequency domain).

We then numerically calculate the RF-pipeline-limited OFO performance. Let $\bm{S}$ be a computer-generated sampling-frequency-limited sawtooth waveform with unit amplitude, ramp frequency $f_{m}$, and sample frequency $f_{s}=9.85\text{ GSPS}$. Writing the E-field of the modulated light in phasor notation and isolating it's dependence on the phase modulation, we can extract the expected performance from the spectrum

\begin{equation}
\mathcal{F}\left\{\exp\left(ia\mathbf{\bm{\theta}}\right)\right\} 
\end{equation}

where

\begin{equation}
\mathbf{\bm{\theta}}=\mathcal{F}^{-1}\left\{ T_{\text{DTO}}\left\{ \mathcal{F}\left\{ \boldsymbol{S}\right\} \right\} \right\}
\end{equation}

and $\mathcal{\mathcal{F}}$ denotes the numerical fast Fourier transform. The value $a$ (modulation amplitude) is varied until the predicted conversion loss is minimized; performance at this amplitude is then recorded. Repeating this procedure for various $f_{m}\in[40,1600]\text{ MHz}$ produces the dashed lines in Figure \ref{fig:OFO-combined-performance}. We find good agreement with measured performance. This suggests that predistorting the AWG waveform to counter the DAC-to-optical transfer function would improve performance.

\subsection{Performance at higher shift indices}

\begin{figure}
    \centering
    \hspace*{-1.5cm}
    \subfloat[Conversion loss for various $N$.]{\includegraphics[width=0.6\textwidth]{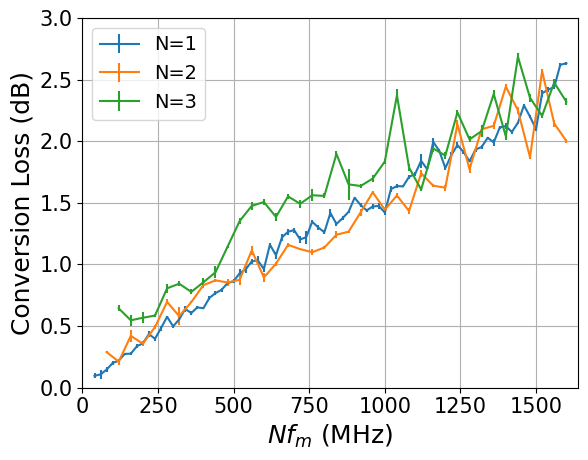}}
    \hfill
    \subfloat[Suppression for various $N$.]{\includegraphics[width=0.6\textwidth]{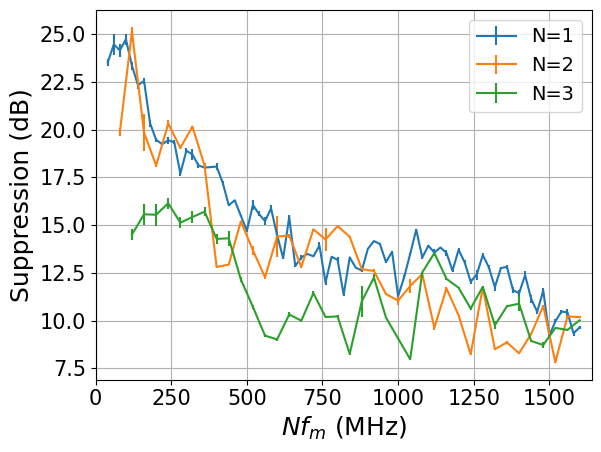}}
    \hspace*{-1.5cm}

    \caption{\label{fig:RFSoC-multi_index-performance} RFSoC performance for shift indices $N=1,2,3$ as a function of the imparted frequency shift $Nf_{m}$.}
\end{figure}

A prospective way to improve performance for a given DAC frequency is to increase $N$ (Eq. \ref{eq:Vserrodyne}). Increasing $N$ yields induced frequency shifts of $Nf_m$ while potentially maintaining comparable performance to the $N=1$ case. Low $V_{\pi}$ is advantageous as the maximum value of $N$ is ultimately limited by the EOM maximum power. Here, it limits us to $N<5$. We measured performance for our setup at these higher shift indices (Figure \ref{fig:RFSoC-multi_index-performance}).

\begin{figure}
    \centering
    \hspace*{-1.5cm}
    \subfloat[Performance for $N=2$.]{\includegraphics[width=0.6\linewidth]{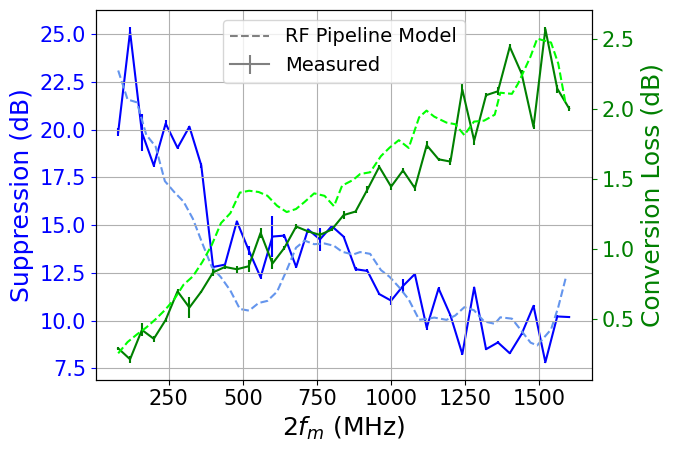}}
    \hfill
    \subfloat[Performance for $N=3$.]{\includegraphics[width=0.6\linewidth]{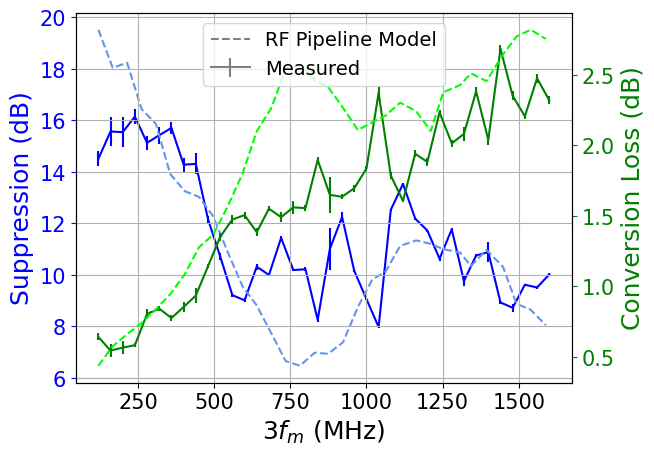}}
    \hspace*{-1.5cm}
    
    \caption{\label{fig:RFSoC-higher-N-performance}RFSoC performance for shift indices $N=2,3$ as a function of the imparted frequency shift $Nf_{m}$.}
\end{figure}

We found no significant gains in performance. We checked for change in the DAC-amplifier transfer function at higher output power but did not see notable change. Furthermore, the curves shown here match simulations of the RF pipeline performance at higher modulation amplitudes (Figure \ref{fig:RFSoC-higher-N-performance}). Our Finite Sampling model suggests that we should see improved performance as a function of induced shift frequency $Nf_m$, thus we suspect that distortions from the amplifier---and to a lesser extent the EOM---significantly degrades performance in going to higher $N$ for a fixed $f_m$. This is further support that performance is limited by imperfections in the DAC-to-EOM RF pipeline.

\section{Application to PDH offset locking}
\begin{figure}[h]
    \centering
    \includegraphics[width=0.6\textwidth]{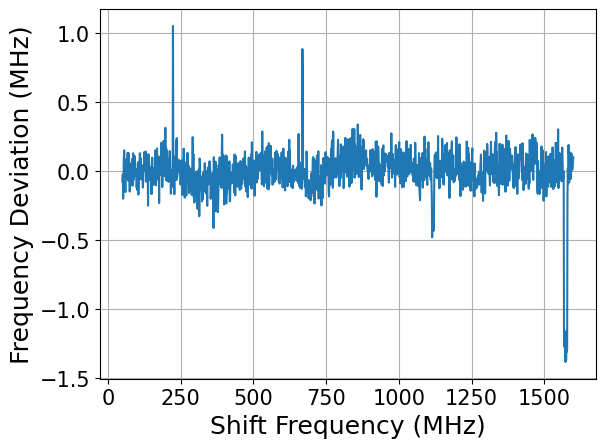}
    \caption{\protect\label{fig:PDH_Sweep}Frequency deviations from a linear sweep of a PDH offset lock swept over $50\text{ MHz}$ to $1600\text{ MHz}$ using the serrodyne apparatus as the frequency shifter. Aside from the 4 large deviations, we are fundamentally limited by the resolution of the wavemeter we are using to measure the lasing frequency.}
\end{figure}

We evaluated the usefulness of our serrodyne shifting apparatus by using it to implement a PDH offset lock sweep \cite{thorpe_laser_2008}. We used an 822 nm ECDL (Toptica DLC Pro), which was split along two beam paths. One path was monitored by a HighFinesse WSU2 Wavemeter, which has a resolution of $200\text{ kHz}$. The other path was serrodyne frequency shifted and we PDH locked to the +1 sideband of the resulting spectrum. The serrodyne modulation frequency was then swept from $50$ to $1600\text{ MHz}$ over the span of 30 seconds while we monitored the lasing frequency on the wavemeter. We then subtracted off a linear fit to the measured frequency v. time to yield the frequency deviations from a perfectly linear sweep, which is shown in Figure \ref{fig:PDH_Sweep}.

We observed anomalous spikes in these deviations, reaching up to $\sim1.5\text{ MHz}$. We tried different sweep periods ranging from 300 seconds down to 10 seconds, but continued to observe these large deviations. Furthermore, the frequency at which these deviations occurred would vary between days. Direct observation of the RFSoC output at each of these frequencies did not reveal any excursions in modulation frequency to the magnitude shown here. Thus these are not features of the serrodyne shifter, but either the wavemeter or our PDH locking setup. Regardless, this demonstrates the viability of the RFSoC as a serrodyne frequency shifter for use in tunable PDH offset locks.

\section{Conclusion}
In summary, we used commercial optoelectronic components to produce continuous wideband ($\sim1\text{ GHz}$) optical frequency offsets in a CW laser with $>10\text{ dB}$ spur suppression, $<2\text{ dB}$ conversion loss and smoothly varying gain (Fig. \ref{fig:OFO-combined-performance}). Relative to other demonstrations we improve on simultaneous optimization of these parameters (Table \ref{tab:Summary-of-OFO}). A simple model accounting for DAC-to-optical transfer function imperfections predicts the observed serrodyne performance. We've confirmed the utility of our OFO by continuously sweeping the offset in a PDH offset lock over the range of $50$ to $1600\text{ MHz}$ without breaking lock.

Going forward, as the cost of high-bandwidth DACs continues to decrease, we anticipate that high-bandwidth, high-suppression OFO will emerge as a routine tool. A natural next step is to apply predistortion to the waveform using a true arbitrary waveform generator.

\section{Back matter}

\textbf{Funding:} The authors acknowledge funding from the Army Research Laboratory (ARL) (Sponsor Award Number W911NF2420107: improved light-matter interfaces for state-control and readout of quantum systems).\\

\noindent\textbf{Disclosures:} The authors declare no conflicts of interest.\\

 \noindent\textbf{Data Availability:} The data that support the findings of this study are available from Roame Hildebrand upon reasonable request.

\section{References}

\begingroup
\renewcommand{\section}[2]{}

\bibliographystyle{aip-modified}
\bibliography{serrodyne,roame_PDH,wance}

@article{johnson_broadband_2010,
	title = {Broadband {Optical} {Serrodyne} {Frequency} {Shifting}},
	volume = {35},
	issn = {0146-9592, 1539-4794},
	url = {http://arxiv.org/abs/0909.1834},
	doi = {10.1364/OL.35.000745},
	number = {5},
	urldate = {2023-12-30},
	journal = {Opt. Lett.},
	author = {Johnson, D. M. S. and Hogan, J. M. and Chiow, S.-w and Kasevich, M. A.},
	month = mar,
	year = {2010},
	note = {arXiv:0909.1834 [physics]},
	pages = {745},
}

@article{houtz_wideband_2009,
	title = {Wideband, {Efficient} {Optical} {Serrodyne} {Frequency} {Shifting} with a {Phase} {Modulator} and a {Nonlinear} {Transmission} {Line}},
	volume = {17},
	url = {https://opg.optica.org/oe/abstract.cfm?URI=oe-17-21-19235},
	doi = {10.1364/OE.17.019235},
	number = {21},
	journal = {Opt. Express},
	author = {Houtz, Rachel and Chan, Cheong and M{\"u}ller, Holger},
	month = oct,
	year = {2009},
	note = {Publisher: Optica Publishing Group},
	pages = {19235--19240},
}

@article{barbiero_broadband_2023,
	title = {Broadband serrodyne phase modulation for optical frequency standards and spectral purity transfer},
	volume = {48},
	url = {https://opg.optica.org/ol/abstract.cfm?URI=ol-48-7-1958},
	doi = {10.1364/OL.485064},
	number = {7},
	journal = {Opt. Lett.},
	author = {Barbiero, M. and Salvatierra, J. P. and Risaro, M. and Clivati, C. and Calonico, D. and Levi, F. and Tarallo, M. G.},
	month = apr,
	year = {2023},
	note = {Publisher: Optica Publishing Group},
	pages = {1958--1961},
}

@article{lee_all-optical_2021,
	title = {All-optical serrodyne frequency shifter},
	volume = {29},
	issn = {1094-4087},
	url = {https://opg.optica.org/abstract.cfm?URI=oe-29-17-26608},
	doi = {10.1364/OE.432242},
	language = {en},
	number = {17},
	urldate = {2023-12-30},
	journal = {Opt. Express},
	author = {Lee, K. F. and Kanter, G. S.},
	month = aug,
	year = {2021},
	pages = {26608},
}

@article{kim_40_2020,
	title = {40 {W}, 780 nm laser system with compensated dual beam splitters for atom interferometry},
	volume = {45},
	issn = {0146-9592, 1539-4794},
	url = {https://opg.optica.org/abstract.cfm?URI=ol-45-23-6555},
	doi = {10.1364/OL.404430},
	language = {en},
	number = {23},
	urldate = {2023-12-30},
	journal = {Opt. Lett.},
	author = {Kim, Minjeong and Notermans, Remy and Overstreet, Chris and Curti, Joseph and Asenbaum, Peter and Kasevich, Mark A.},
	month = dec,
	year = {2020},
	pages = {6555},
}

@article{zhu_spectral_2022,
	title = {Spectral control of nonclassical light pulses using an integrated thin-film lithium niobate modulator},
	volume = {11},
	issn = {2047-7538},
	url = {https://www.nature.com/articles/s41377-022-01029-7},
	doi = {10.1038/s41377-022-01029-7},
	language = {en},
	number = {1},
	urldate = {2024-02-13},
	journal = {Light Sci Appl},
	author = {Zhu, Di and Chen, Changchen and Yu, Mengjie and Shao, Linbo and Hu, Yaowen and Xin, C. J. and Yeh, Matthew and Ghosh, Soumya and He, Lingyan and Reimer, Christian and Sinclair, Neil and Wong, Franco N. C. and Zhang, Mian and Lon{\v c}ar, Marko},
	month = nov,
	year = {2022},
	pages = {327},
}

@article{thorpe_laser_2008,
	title = {Laser frequency stabilization and control through offset sideband locking to optical cavities},
	volume = {16},
	issn = {1094-4087},
	url = {https://opg.optica.org/oe/abstract.cfm?uri=oe-16-20-15980},
	doi = {10.1364/OE.16.015980},
	language = {en},
	number = {20},
	urldate = {2024-02-13},
	journal = {Opt. Express},
	author = {Thorpe, J. I. and Numata, K. and Livas, J.},
	month = sep,
	year = {2008},
	pages = {15980},
}

@article{kawanishi_high-speed_2007,
	title = {High-{Speed} {Control} of {Lightwave} {Amplitude}, {Phase}, and {Frequency} by {Use} of {Electrooptic} {Effect}},
	volume = {13},
	issn = {1558-4542},
	url = {https://ieeexplore.ieee.org/document/4084539},
	doi = {10.1109/JSTQE.2006.889044},
	number = {1},
	urldate = {2024-03-14},
	journal = {IEEE Journal of Selected Topics in Quantum Electronics},
	author = {Kawanishi, Tetsuya and Sakamoto, Takahide and Izutsu, Masayuki},
	month = jan,
	year = {2007},
	note = {Conference Name: IEEE Journal of Selected Topics in Quantum Electronics},
	pages = {79--91},
}

@article{uttam_precision_1985,
	title = {Precision time domain reflectometry in optical fiber systems using a frequency modulated continuous wave ranging technique},
	volume = {3},
	issn = {0733-8724},
	url = {http://ieeexplore.ieee.org/document/1074315/},
	doi = {10.1109/JLT.1985.1074315},
	number = {5},
	urldate = {2024-03-14},
	journal = {J. Lightwave Technol.},
	author = {Uttam, D. and Culshaw, B.},
	year = {1985},
	pages = {971--977},
}

@article{macdonald_frequency_1981,
	title = {Frequency domain optical reflectometer},
	volume = {20},
	issn = {0003-6935, 1539-4522},
	url = {https://opg.optica.org/abstract.cfm?URI=ao-20-10-1840},
	doi = {10.1364/AO.20.001840},
	language = {en},
	number = {10},
	urldate = {2024-03-14},
	journal = {Appl. Opt.},
	author = {MacDonald, R. I.},
	month = may,
	year = {1981},
	pages = {1840},
}

@article{zhou_laser_2020,
	title = {Laser frequency shift up to 5 {GHz} with a high-efficiency 12-pass 350-{MHz} acousto-optic modulator},
	volume = {91},
	issn = {0034-6748, 1089-7623},
	url = {https://pubs.aip.org/rsi/article/91/3/033201/1030780/Laser-frequency-shift-up-to-5-GHz-with-a-high},
	doi = {10.1063/1.5142314},
	language = {en},
	number = {3},
	urldate = {2024-03-16},
	journal = {Review of Scientific Instruments},
	author = {Zhou, Chao and He, Chuan and Yan, Si-Tong and Ji, Yu-Hang and Zhou, Lin and Wang, Jin and Zhan, Ming-Sheng},
	month = mar,
	year = {2020},
	pages = {033201},
}

@article{johnson_serrodyne_1988,
	title = {Serrodyne optical frequency translation with high sideband suppression},
	volume = {6},
	issn = {07338724},
	url = {http://ieeexplore.ieee.org/document/3974/},
	doi = {10.1109/50.3974},
	number = {1},
	urldate = {2024-03-17},
	journal = {J. Lightwave Technol.},
	author = {Johnson, L.M. and Cox, C.H.},
	month = jan,
	year = {1988},
	pages = {109--112},
}

@article{donley_double-pass_2005,
	title = {Double-pass acousto-optic modulator system},
	volume = {76},
	issn = {0034-6748, 1089-7623},
	url = {https://pubs.aip.org/rsi/article/76/6/063112/353058/Double-pass-acousto-optic-modulator-system},
	doi = {10.1063/1.1930095},
	language = {en},
	number = {6},
	urldate = {2024-04-16},
	journal = {Review of Scientific Instruments},
	author = {Donley, E. A. and Heavner, T. P. and Levi, F. and Tataw, M. O. and Jefferts, S. R.},
	month = jun,
	year = {2005},
	pages = {063112},
}

@article{kodigala_high-performance_2024,
	title = {High-performance silicon photonic single-sideband modulators for cold-atom interferometry},
	volume = {10},
	issn = {2375-2548},
	url = {https://www.science.org/doi/10.1126/sciadv.ade4454},
	doi = {10.1126/sciadv.ade4454},
	language = {en},
	number = {28},
	urldate = {2024-12-06},
	journal = {Sci. Adv.},
	author = {Kodigala, Ashok and Gehl, Michael and Hoth, Gregory W. and Lee, Jongmin and DeRose, Christopher T. and Pomerene, Andrew and Dallo, Christina and Trotter, Douglas and Starbuck, Andrew L. and Biedermann, Grant and Schwindt, Peter D. D. and Lentine, Anthony L.},
	month = jul,
	year = {2024},
	pages = {eade4454},
}

@misc{tu_quadrature_2024,
	title = {Quadrature amplitude modulation for electronic sideband {Pound}-{Drever}-{Hall} locking},
	copyright = {Creative Commons Attribution 4.0 International},
	url = {https://arxiv.org/abs/2409.08764},
	doi = {10.48550/ARXIV.2409.08764},
	urldate = {2025-04-10},
	publisher = {arXiv},
	author = {Tu, J. and Restelli, A. and Tsui, T. -C. and Weber, K. and Spielman, I. B. and Rolston, S. L. and Porto, J. V. and Subhankar, S.},
	year = {2024},
	note = {Version Number: 1},
}

@article{lo_precise_2017,
	title = {Precise tuning of single-photon frequency using an optical single sideband modulator},
	volume = {4},
	issn = {2334-2536},
	url = {https://opg.optica.org/abstract.cfm?URI=optica-4-8-919},
	doi = {10.1364/OPTICA.4.000919},
	language = {en},
	number = {8},
	urldate = {2025-04-10},
	journal = {Optica},
	author = {Lo, Hsin-Pin and Takesue, Hiroki},
	month = aug,
	year = {2017},
	pages = {919},
}

@article{hu_-chip_2021,
	title = {On-chip electro-optic frequency shifters and beam splitters},
	volume = {599},
	issn = {0028-0836, 1476-4687},
	url = {https://www.nature.com/articles/s41586-021-03999-x},
	doi = {10.1038/s41586-021-03999-x},
	language = {en},
	number = {7886},
	urldate = {2025-04-10},
	journal = {Nature},
	author = {Hu, Yaowen and Yu, Mengjie and Zhu, Di and Sinclair, Neil and Shams-Ansari, Amirhassan and Shao, Linbo and Holzgrafe, Jeffrey and Puma, Eric and Zhang, Mian and Lon{\v c}ar, Marko},
	month = nov,
	year = {2021},
	pages = {587--593},
}

@article{drever_laser_1983,
	title = {Laser phase and frequency stabilization using an optical resonator},
	volume = {31},
	copyright = {http://www.springer.com/tdm},
	issn = {0721-7269, 1432-0649},
	url = {http://link.springer.com/10.1007/BF00702605},
	doi = {10.1007/bf00702605},
	language = {en},
	number = {2},
	urldate = {2025-07-15},
	journal = {Appl. Phys. B},
	author = {Drever, R. W. P. and Hall, J. L. and Kowalski, F. V. and Hough, J. and Ford, G. M. and Munley, A. J. and Ward, H.},
	month = jun,
	year = {1983},
	note = {Publisher: Springer Science and Business Media LLC},
	pages = {97--105},
}

@article{kapoor_electro-optic_2025,
	title = {Electro-optic frequency shift of single photons from a quantum dot},
	volume = {14},
	copyright = {http://creativecommons.org/licenses/by/4.0},
	issn = {2192-8614},
	url = {https://www.degruyterbrill.com/document/doi/10.1515/nanoph-2024-0550/html},
	doi = {10.1515/nanoph-2024-0550},
	language = {en},
	number = {11},
	urldate = {2025-07-17},
	journal = {Nanophotonics},
	author = {Kapoor, Sanjay and Rodek, Aleksander and Miko{\l }ajczyk, Micha{\l } and Szuniewicz, Jerzy and So{\'s}nicki, Filip and Kazimierczuk, Tomasz and Kossacki, Piotr and Karpi{\'n}ski, Micha{\l }},
	month = jun,
	year = {2025},
	note = {Publisher: Walter de Gruyter GmbH},
	pages = {1775--1782},
}

\end{document}